\documentstyle{article}
\def\case#1/#2{{\textstyle {#1\over #2}}}
\def\re{\par\noindent}

\begin{document}

\title{EXOTIC BLACK HOLES\footnote{Talk presented at the 8th
Nishinomiya--Yukawa Memorial Symposium, October 1993.}}

\author{ Ian G. Moss \\
{\it  Department of Physics, University of Newcastle--upon--Tyne,
NE1 7RU, U.K.} }

\maketitle

\section*{Abstract}

Black hole solutions can be used to shed light on general issues in
General Relativity and Quantum Physics. Black--hole hair, entropy
and naked singularities are considered here, along with some
implications for the inflationary universe scenario.

\section{Introduction}

Black holes form as a result of gravitational collapse which may be
very variable in nature, but a significant simplification results from
the observation that the final state of gravitational collapse seems to
be characterised by only a few parameters. This is the phenomenon that
John Wheeler described as the loss of hair by the black hole (Ruffini
and Wheeler 1971) . It seems to be a feature of the surface or the
event horizon that defines the limit of communication with the
outside.

Over the last few years a number of new black hole solutions have been
discovered. These include solutions that do not lose their hair
entirely, they have some residual fields outside of their event
horizons and sometimes need extra parameters to distinguish them. By
studying these solutions it is possible to understand fundamental
issues of black hole physics a little better.

One important issue is that of black hole thermodynamics. For an event
horizon of area ${\cal A}$ we are able to associate an entropy S,
\begin{equation}
S=\case1/4 {\cal A}
\end{equation}
(in units where $\hbar=c=G=k=1$). It seems unusual to assign an entropy
to a classical field configuration, but the justification lies in the
extent to which the laws of thermodynamics can be applied to a black
hole. The relationship is consistent with the second law of
thermodynamics and seems natural on the grounds of information
swallowed by the hole.

The discovery made by Stephen Hawking was that thermodynamical
equilibrium of the hole and its surroundings could only be achieved by
including quantum mechanical pair creation. This allows the hole to
radiate as well as absorb energy. The temperature is given by the
Hawking formula $T=\kappa/2\pi$, where $\kappa$ is the surface
gravity of the event horizon.

Even this, however, cannot take us fully into the range of
thermodynamics because the equilibrium between a black hole and its
surroundings would usually be unstable. This is an effect of having a
negative specific heat. The black hole forces us to reconsider the
basic laws of thermodynamics just as is forces us to question the basis
of unitary evolution in quantum mechanics.

There is also an interesting class of black hole solutions that
include a cosmological constant. The loss of hair for these
holes is important for the homogeneity of the early universe
if it underwent a period of inflation. In fact the existence of these
solutions shows explicitly that the universe does not become
homogeneous, but it is consistent with the `no hair' theorems for
the geometry to approach de Sitter assymptotically.

Beyond the event horizon, in the interior, lies a region of space that
pushes our physical knowledge to its limits. According to general
relativity there is a spacelike singularity, or a null singularity if
the black hole is rotating. This singularity would define the end of
time as far as classical evolution is concerned.

A far more serious condition is encountered if the singularity is
timelike. The class of solutions that describe black holes in a closed
universe with a cosmological constant do seem to have a singularities
of this kind. The time evolution from a singularity is not unique in
Einstein gravity and therefore the spacetime becomes unpredictable
before time comes to an end. It is even possible for time--like curves
to loop back on themselves. These features demand a theory which goes
beyond Einstein's theory of gravity, possibly a theory of quantum
gravity.

\section{No--Hair Theorems}

Originally the no--hair theorems arose from attempts to understand the
nature of the event horizon that remains after gravitational collapse
(Israel 1967).  The early work is summarised by Price's theorem (Price
1972):
\medskip

The only stationary solutions to the massless wave equations with spin
$s=0,1,2$ on a spherically symmetrical black hole background have
angular momentum less than $s$.
\medskip

The implication is that all of the other modes are radiated away during
the collapse and that the event horizon is quite easy to characterise.
Any field which violates the conclusion of Price's theorem might
therefore be described as black hole hair.

There is a more useful view, which can be found in the work of Brandon
Carter for example (Carter 1979), that the no--hair theorems are
statements about the number of  parameters that are needed to specify
the equilibrium state. In Einstein--Maxwell theory there are just three
of these parameters:
\medskip

{\parindent=0pt
mass $M$\hfill surface gravity $\kappa$

charge $Q$\hfill potential $\alpha$

angular momentum $J$\hfill rotation $\Omega$

}
\medskip

In this list the set of quantities that are measured far from the hole
are listed on the left, but there is also an alternative set of
parameters that characterise the horizon itself listed on the right.
These quantities are defined by properties of the various killing
fields that are present.

For equilibrium the solutions have a killing field $k$ which is
timelike at large distances from the hole
but also, because of a theorem due to Hawking (Hawking 1972), a killing
field $m$ for rotational symmetry. The horizon is generated by a null
combination of the two fields $l=k+\Omega m$. This defines the rotation
rate $\Omega$ and the surface gravity is then defined by
$\nabla\,(l^2)=\kappa\,l$. Finally, the potential $\alpha$ is the
ordinary electrostatic potential at the horizon.

The new black hole solutions that have been discovered in the past few
years have additional scalar or gauge fields that are not present in
the case of electromagnetism. As the table below shows, we include as
solutions cases that are unstable and even ones that have the `wrong'
spacetime signature. The reason for such a liberal approach is that
black hole thermodynamics is a quantum phenomenon and the solutions may
represent virtual processes.
\medskip

Table 1. Black hole solutions with conventional gravity and a variety
of other fields.
\medskip

\begin{tabular}{lllll} \hline
type&fields&matter Lagrangian&number&stable\\ \hline
dilaton&dilaton&
$-\case1/4e^{-2\phi}F^2-\case1/2(\nabla\phi)^2$&1&Yes\\
skyrmion&chiral&
$-\case1/4k^2{\rm tr}A^2-\case1/{32}g^2{\rm tr}(A\wedge A)^2$&N&Yes\\
sphaleron&gauge&$-\case1/4{\rm tr}F^2$&N&No\\
monopole&gauge/higgs&
$-\case1/4{\rm tr}F^2-(D\phi)^\dagger(D\phi)-V(\phi)$&N&Yes\\
vortex&gauge/higgs&
$-\case1/4{\rm tr}F^2-(D\phi)^\dagger(D\phi)-V(\phi)$&N&N/A\\
\hline
\end{tabular}
\bigskip
\bigskip

\noindent{\it Dilaton} (Gibbons and Maeda 1988)
\smallskip

The dilaton field appears in theories in which the spacetime dimension
starts out larger than four. Special examples result from string
theories.
The charged black hole solutions with this matter action have
non--constant scalar field values outside the event horizon. However
the solutions are unique, provided that the combination $Qe^{-\phi}$ is
interpreted as the charge. These solutions therefore only have hair in
the weak sense of having non--trivial fields outside the horizon.
\medskip

\noindent{\it Skyrmion} (Luckock 1987)
\smallskip

These solutions represent a black hole immersed inside
a cloud of pions. In the Skyrmion model pions are represented by
$SU(2)$ matrices, $U=\exp(iw(r)\sigma\cdot x)$. The Lagrangian in table
1 is given in terms of  $A=g^{-1}U^{-1}d\,U$ with constants $k$ and
$g$. The topological solitons of this theory represent baryons, with
the baryon number $N$ given by the difference between $w$ at infinity
and at the origin, when there is no horizon. Black hole solutions
have a horizon at $r=r_h$, and the value
of $w(r_h)$ appears as two branches--an unstable and a stable
branch--connected to each value of $N$ (Luckock 1986, Droz 1991).
\medskip

\noindent{\it Sphaleron} (Volkov and Gal'tsov 1989)
\smallskip

These are solutions with a Yang--Mills gauge field, typically with the
gauge group $SU(2)$ (Volkov 1989, Bizon 1990, K\"unzle 1990). The
fields look like
\begin{equation}
A={1-w(r)\over 2g}U^{-1}d\,U,\quad U=e^{i\sigma\cdot x},
\end{equation}
with $w(r)\to\pm1$ as $r\to\infty$. The solutions are unstable
(Straumann 1990) but lie midway between topologically distinct vacua
and can represent tunnelling.
\medskip

\noindent{\it Monopole} (Breitenlohner, Forg\'acs and Maison 1992)
\smallskip

With a Higgs field as well as the gauge field there are a class of
monopole solutions in which the Higgs field does not take the vacuum
value outside of the event horizon (Breitenlohner 1992, Lee 1992). The
gauge fields take the same form as for the sphaleron, but with
$w(r)\to0$ as $r\to\infty$, and the Higgs doublet field is of the
hedgehog form  $\phi=\chi(r)x\cdot\sigma/r$.
\medskip

\noindent{\it Vortex} (Dowker, Gregory and Traschen 1992)
\smallskip

These solutions with a Yang--Mills gauge field and Higgs field are
unlike the rest in that they have Riemannian spacetime signature
$(++++)$. The reason for including these solutions is entirely due to
the quantum nature of black--hole thermodynamics. For this reason these
solutions are often described as examples of `quantum hair' (Coleman
1991). The gauge and Higgs fields for a $U(1)$ gauge group are
\begin{equation}
A=w(r)dt,\quad\phi=\chi(r)e^{ig\gamma t}.
\end{equation}
For a Higgs charge $ng$ and fermion charge $g$, $w\to\gamma/n$ as
$r\to\infty$. The solutions are topologically non--trivial only if
$n>0$, a rather `non--standard' charge for the Higgs field.
There is a winding number $0\le N<n$ for the Higgs phase factor,
i.e $\gamma=2\pi N/g\beta$.

\medskip
\section{Black Hole Thermodynamics}

The remarkable extension of thermodynamics to black holes is based upon
the laws of classical black hole mechanics (Bardeen 1973):
\medskip

(0) The event horizon is described by a quantity $\kappa$, the surface
gravity.

(1) Under external influences,
\begin{equation}
\delta M=\case1/4(\kappa/2\pi)\delta {\cal A}+\hbox{work terms}.
\end{equation}

(2) The area increase law: $\delta {\cal A}\ge1$.

(3) The limit $\kappa=0$ cannot be reached in a finite time.

\medskip
These laws are dependent upon a number of topological assumptions and
energy conditions, e.g. the weak energy condition $T_{ab}l^al^b\ge0$
for null vectors $l$. Until recently the first law had only been
investigated for particular models, and this is even more so for the
third law (Israel 1992).

The main thermodynamic result missing from the classical laws
is the statement of thermal equilibrium. This is supplied by
Hawking's famous result,

\medskip
(0) A black hole can be in thermal equilibrium with radiation at the
Hawking temperature.

\medskip
In order to establish this result it is necessary to look at quantum
field theory on a black hole background. Underlying the result is the
periodicity of black hole solutions in imaginary time which is
analagous to the behaviour of thermally averaged Green functions. The
period $\beta=2\pi/\kappa$ is fixed by regularity of the metric at the
event horizon.

The first law enables us to relate the entropy of a black hole to the
area of the event horizon. It is important to establish whether the
first law is still valid for the black hole solutions that are
mentioned in table 1, and the recent work of Bob Wald goes some way
towards this end (Wald 1993). It should also be born in mind that the
equilibrium is usually unstable and only a subset of thermodynamics is
strictly valid. Some discussion of this issue and attempts towards a
better theory of black--hole thermodynamics can be found in the work of
York (Brown 1991).

Another way to calculate the entropy is to employ the partition
function, for example for the grand canonical ensemble. Calculation of
the entropy $S(\beta,\mu)$ then proceeds as follows:

(1) Define the partition function
\begin{equation}
Z={\rm tr}(e^{-\beta M-\mu Q})
\end{equation}
by a path integral over field configurations with Riemannian geometry,
period $\beta$ and potential $\mu$.

(2) Expand the partition function about a solution to the classical
equations,
\begin{equation}
Z\sim(\det\Delta)^{-1/2}e^{-I(\beta,\mu)}.
\end{equation}
The determinant contains the contribution from the radiation gas
and the classical black--hole action $I$ represents the coherent
fields (Hawking 1979).

(3) Use a canonical decomposition of the action to show that
for Einstein gravity plus matter (Moss 1992),
\begin{equation}
I=\beta M-\case1/4{\cal A}-\beta\Omega J+\beta\,\alpha_h Q_h
-\beta\,\alpha_\infty Q_\infty\label{action}
\end{equation}
All of the quantities in this expression are measured at the horizon or
at infinity. The gauge potential plus a topological term appears as
$\alpha$.

(4) Read off the thermodynamic potential $W=-T\ln Z$, and compare the
result with
\begin{equation}
W=M-\beta^{-1}S-\sum\mu_iN_i
\end{equation}
to deduce that the contribution to the action from the black hole is
$S=\case1/4 {\cal A}.$ (If there is more than one black hole solution
then the total entropy would include contributions from each one).

It is step (3) that contains most of the technical calculation. We want
a formula for the action of a black hole that only depends on local
terms at the horizon and infinity. We use the basic property of black
hole solutions that they have a `timelike' killing vector $k$, a
spacelike killing vector $m$ and an event horizon ${\cal S}_h$.

Begin by constructing spaces $\Sigma$ of constant time which stretch
from the event horizon to infinity and pick a tetrad. The coefficients
of this tetrad in a coordinate basis define the lapse and shift
functions $N$ and $N^i$,
\begin{equation}
\omega^0=N\,dt,\quad
\omega^m=\omega_i^{\ m}(N^idt+dx^i).
\end{equation}
The surface metric $h_{ij}=\omega_i^{\ m}\omega_j^{\ m}$ summed over
$m$.

Take as an example the matter Lagrangian density for scalar
electrodynamics, with scalar field $\phi$ and gauge potential $A_\nu$.
The momentum conjugate to the spatial components of the gauge potential
is the electric field density $E^i$. The Lagrangian has a canonical
decomposition into
\begin{equation}
{\cal L}_mN\sqrt{h}=
\pi\dot\phi^\dagger+\pi^\dagger\dot\phi+E^i\dot A_i
-A_t(E^i_{\ ,i}-\rho)-N{\cal H}-N^i{\cal H}_i
+(E^i A_t)_{,i}\label{lagr}.
\end{equation}
The boundary terms is this expression are crucial.

The charge density $\rho=-i(\pi^\dagger\phi-\phi^\dagger\pi)$, and the
magnetic field  $B=h^{1/2}\nabla\wedge A$. The contributions to the
Hamiltonian from each of the fields are tabulated below.
\bigskip

Table 2. Hamiltonian and momentum densities for various fields.
\vspace{1pc}

\begin{tabular}{lll} \hline
fields&${\cal H}$&${\cal H}_i$\\ \hline
scalar&
$h^{-1/2}\pi\pi^\dagger+h^{1/2}\nabla\phi^\dagger\cdot \nabla\phi$&
$\pi^\dagger\phi_{,i}+\pi{\phi_{,i}}^\dagger$\\
vector&
$\case1/2h^{-1/2}\left(E\cdot E+B\cdot B\right)$&
$h^{1/2}E\wedge B$\\
dilatonic&
$\case1/2h^{-1/2}\left(E\cdot D+B\cdot H\right)$&
$h^{1/2}E\wedge H$\\
\hline
\end{tabular}
\bigskip

Time translation symmetry {\it up to a gauge transformation} implies
that
$\dot\phi=ig\gamma\phi$ and $\quad\dot A_i=\gamma_{,i}$. Set
$\alpha=\gamma+A_t$. The matter action coming from integrating equation
\ref{lagr} becomes simply
\begin{equation}
I_m=\beta \int d\mu(\Sigma)(N{\cal H}+N^i{\cal H}_i)
+\beta\,\alpha_h Q_h-\beta\,\alpha_\infty Q_\infty.
\end{equation}
In the vortex solution, for example, the charge is a constant and
$Q_h=Q_\infty$. At the horizon, $A_t=0$ and therefore
$\alpha_h=\gamma$, whilst at infinity $\alpha_\infty=0$.

Canonical decomposition of the Einstein action can be used to derive a
similar result for the gravitational part of the action. The details
will be omitted because it is also possible obtain the result more
easily from killing vector identities (Gibbons 1977,Moss 1992),
\begin{equation}
I_g=\beta M-\case1/4 {\cal A}-\Omega J -\beta
\int d\mu(\Sigma)(N{\cal H}+N^i{\cal H}_i).
\end{equation}
The last term in this part of the action cancels with the identical
term in the matter action to leave the result for the total action
\ref{action} that was quoted above.

The result extends to any matter field which has a canonical
decomposition, and the dilaton theory is an example. If we set
\begin{equation}
D=e^{-2\phi}E\quad B=e^{2\phi}H,
\end{equation}
then Maxwell field part of the Hamiltonian takes the form given in the
table.
The Maxwell equations take the natural form for electrodynamics of
continuous media, $\nabla\cdot D=\rho$ etc., together with the dilaton
field equation $\nabla^2\phi={\cal H}\sqrt{h}$.

Fundamental departures between the entropy and area relationship do
occur when the gravitational part of the theory is modified. One
example is to include terms in the Lagrangian that have two or more
powers of the curvature tensor (Jacobson 1993, Visser 1993). In these
cases the entropy is modified,
\begin{equation}
S=\case1/4{\cal A}+4\pi\int d\mu({\cal S}_h)
{\partial {\cal L}^{extra}\over\partial R_{\mu\nu\rho\sigma}}
g^\perp_{\mu\rho}g^\perp_{\nu\sigma},
\end{equation}
where $g^\perp_{\mu\rho}$ is the projection of the metric orthogonal to
the horizon.

\section{Cosmic Censorship}

Can we travel into a black hole and through into another universe,
seeing the singularity, closed timelike curves etc. on the way?
For ordinary black
holes, surrounded by empty space, it seems that a singularity forms
at the Cauchy horizon which prevents passage throught the hole (Israel
1992). This is caused by radiation
from the exterior, which becomes infinitely blue--shifted near to the
horizon resulting in a diverging energy density.
\begin{figure}[t]
   \vspace{5cm}
   \caption{These Penrose diagrams show different possibilities where
   the singularity can be invisible or visible}
\end{figure}

The situation with a cosmological constant is different, as shown in
the figure. Light rays close to the Cauchy horizon have travelled from
a cosmological horizon and not from infinity. The difference in surface
gravities of the two horizons determines whether the wavelength of the
light rays gets red or blueshifted, with redshift and Cauchy horizon
stability when the cosmological horizon has the larger surface gravity
(Mellor 1990, Brady 1992).

Both charged and rotating black hole solutions can be generalised
to include a cosmological constant $\Lambda$ (Carter 1973). The metric
for a rotating hole is quite complicated but simplifies a little
because of a symmetry between the radial coordinate $r$ and colatitude
coordinate $\mu=a\cos\theta$,
\begin{equation}
ds^2=\rho^2(\Delta_r^{-1}dr^2+\Delta_\mu^{-1}d\mu^2)
+\rho^{-2}\Delta_\mu\,\omega^1\otimes\omega^1
-\rho^{-2}\Delta_r\,\omega^2\otimes\omega^2,
\end{equation}
with one--forms,
\begin{equation}
\chi^2\omega^1=dt-a^{-1}\sigma_r^2d\phi\hbox{,~~~~}
\chi^2\omega^2=dt-a^{-1}\sigma_\mu^2 d\phi.
\end{equation}
The metric solution is parameterised as follows,
\begin{eqnarray}
&&\sigma_r=(a^2+r^2)^{1/2}\hbox{,~~~~}
\Delta_r=(a^2+r^2)(1-\case1/3\Lambda r^2)-2Mr\\
&&\sigma_\mu=(a^2-\mu^2)^{1/2}\hbox{,~~~~}
\Delta_\mu=(a^2-\mu^2)(1+\case1/3\Lambda \mu^2)
\end{eqnarray}
and
\begin{equation}
\rho^2=r^2+\mu^2\hbox{,~~~~}
\chi^2=1+\case1/3\Lambda a^2.
\end{equation}
The fully extended spacetime has many asymptotic regions and the
extremal
sections are infinite chains of three--spheres with black holes at
their antipodes. In a situation where a rotating body collapsed to form
a black hole, part of the Penrose diagram would represent the spacetime
outside of the collapsing matter.

\begin{figure}[t]
   \vspace{8cm}
   \caption{Penrose diagram of Kerr--de Sitter ($\theta=\pi/2$).}
\end{figure}

There are three basic types of horizon where the function $\Delta_r$
vanishes:

\begin{tabbing}
${\cal S}_1$\qquad cosmological horizon\\
${\cal S}_2$\qquad black hole event horizon\\
${\cal S}_3$\qquad black hole Cauchy horizon\\
\end{tabbing}

Perturbation theory can be used to study stability of the spacetime
(Chambers 1993). The analogue of the Starobinsky--Teukolsky equations
for the radial modes ${\cal R}$ of a field with spin $s$, frequency
$\omega$ and angular quantum number $m$ are
\begin{equation}
\left({\cal D}_{-s/2}\Delta_r{\cal
D}_{s/2}^\dagger+2(2s-1)i\omega\chi^2\right){\cal R}
=(\nu+s)(\nu-s+1) {\cal R}.
\end{equation}
The gravitational case includes an additional cosmological term,
\begin{equation}
\left({\cal D}_{-1}\Delta_r{\cal D}_1^\dagger+6i\omega\chi^2
-2\Lambda r^2\right){\cal R}=(\nu+2)(\nu-1) {\cal R},
\end{equation}
with radial derivatives
\begin{equation}
{\cal D}_n={\partial\over\partial r}+{iK_r\over \Delta_r}
+n{\Delta_r'\over\Delta_r}\hbox{,~~~~}
K_r=\chi^2(am-\sigma_r^2\omega).
\end{equation}
The equations for the angular modes, which determine the eigenvalues
$(\nu+s)(\nu-s+1)$, are obtained by replacing $r$ by $i\mu$ in the radial
equations. These equations reduce the stability analysis to a
series of scattering problems, but the most important feature is the
combined blue--shift of the ingoing radiation, $e^{\kappa_3v}$ in
the interior and $e^{-\kappa_1v}$ outside, where $v$ is a
coordinate that becomes infinite at the Cauchy horizon. The Cauchy
horizon is stable when the product is finite, i.e. $\kappa_3<\kappa_1$.

The stability region is shown in the figure above, parameterised by the
mass $M$ and rotation parameter $a$ scaled by the cosmological
constant. The range of parameters for which there are three horizons
lies inside the triangular region. On the outer borders two of the
radii are equal, $r_2=r_3$ on OA and $r_1=r_2$ on AC, but the
spacetimes can still be defined. Corresponding surface gravities also
vanish there. Interior lines denote the cases where $\kappa_1=\kappa_3$
(the lower line OA) and $\kappa_1=\kappa_2$ (along OB). The narrow area
between the two lines OA is the stability region.

Although the region is tiny for small values of the cosmological
constant it would be possible to force the parameters of the hole in
the right direction by increasing the rotation rate. If the third law
of black hole mechanics is valid then there is an upper limit to the
rotation rate at the upper line OA in the figure.

\begin{figure}[t]
   \vspace{8cm}
   \caption{The parameter space of the rotating black hole.}
\end{figure}

In fact there is an interesting violation of the third law of black
hole mechanics for de Sitter black holes (Brill 1993). This is easiest
to see for the charged case. The line corresponding to OB in the figure
is given in the charged case by the condition $Q=M$ (Mellor 1989). The
usual centrifugal barrier around $r=0$ is absent for particles which
have equal charge and mass. The singularity is no longer `repulsive'.
If a shell of this type of matter is droped into a black hole, with
parameters close to point B, the mass can be increased beyond B (still
with $Q=M$) and the horizon stripped from the hole. This is a violation
of the third law because $\kappa_2=0$ at B.

The modification of a charged black hole that would lead to stability
involves an increase in the charge to mass ratio and therefore the
validity of the third law for this case is an open issue. It remains to
be seen whether Cauchy horizon stability is a `practical' proposition
in a universe which has a very small cosmological constant.

Finally, there is the possibility that the cosmological constant was
once very large and responsible for a period of inflationary expansion.
The issues of interest in this situation are the stability of the
whole universe rather than just the Cauchy horizon and the existance
of `no hair' theorems that generalise the results for black holes
in empty space (Hawking 1982).

The generalisation of Price's theorem to de Sitter space takes the form:

\medskip
The only stationary solutions to the massless wave equations with spin
$s=0,1,2$ on a spherically symmetrical black hole background in de
Sitter space have angular momentum less than $s$.
\smallskip

In fact the modified Starobinsky--Teukolsky equations given above
have no stationary solutions at all when the rotation parameter $a=0$.
Take the equation,
\begin{equation}
\left({\cal D}_{-s/2}\Delta_r{\cal D}_{s/2}^\dagger
-(\nu+s)(\nu-s+1)\right){\cal R}=0.
\end{equation}
Multiply by the complex conjugate mode solution and integrate
between the horizons $r_1$ and $r_2$,
\begin{equation}
\int_{r_2}^{r_1}dr{\cal R}^\dagger
\left({\cal D}_{-s/2}\Delta_r{\cal D}_{s/2}^\dagger
-(\nu+s)(\nu-s+1)\right){\cal R}=0.
\end{equation}
For physically reasonable solutions it would be necessary for both
${\cal R}$ and ${\cal D}_0{\cal R}$ to be regular at the horizons.
For the case $a=0$, integration by parts gives
\begin{equation}
\int_{r_2}^{r_1}dr
\left(-\Delta_r^{(1-s)} (\Delta_r^{s/2} {\cal R})^{\dagger\prime}
(\Delta_r^{s/2} {\cal R})'
-(\nu+s)(\nu-s+1){\cal R}^\dagger{\cal R}\right)=0.
\end{equation}
The function $\Delta_r$ is positive in this range of integration and
the value of $\nu$ is a non--negative integer when $a=0$.

The only consistent solutions to the Starobinsky--Teukolsky equations
with $\nu\ge s$ are therefore ${\cal R}=0$. Some field components are
not fully determined by solutions to the Starobinsky--Teukolsky
equations and have to be considered seperately (Chambers 1993).
They are only determined up to a solution of
\begin{equation}
{\cal D}_0(r^{1+s}\phi)=0
\end{equation}
and they have angular momentum less than $s$.
(For those familiar with Newman--Penrose formalism, these fields are
the Maxwell scalar $\phi_1$ or the perturbation of the Weyl tensor
component $\Psi_2$.) When $s=1$, the complex scalar is composed of
the radial electric and magnetic fields and the two real constants
of integration are the electric and magnetic charges of the black
hole. When $s=2$, $\phi$ is the perturbation to
the Weyl tensor and the constants of integration are the mass
perturbation and the angular momentum.

The black hole solutions show very explicitly that a period of inflation
need not lead to a universe that is fully homogeneous. On the other
hand, any observer who does not fall into the singularity sees a
local spacetime that comes to look exactly like de Sitter space
and becomes increasingly isotropic. Furthermore, most of the spacetime
behaves this way and it seems that inflation will still work even if
the universe began with large inhomogeneities (Shiromizu 1993).

The decay of perturbations can be seen explicilty in the case
of de Sitter space $a=M=0$ (Mellor 1990).  The modified
Starobinsky--Teukolsky equations can be written as a scattering
problem using the coordinate $r^*$, where $r=\alpha\tanh(r^*/\alpha)$
and $\alpha^2=3/\Lambda$. Then the perturbation equations are
simply
\begin{equation}
\left({d^2\over dr^{*2}}+V(r)\right)\,r{\cal R}=0
\end{equation}
where
\begin{equation}
V(r)=\left(\omega+{is\over r}\right)^2
-(\nu(\nu+s+1)+s(s-1)){\Delta_r\over r^4}.
\end{equation}

The solutions can be written in terms of Jacobi polynomials,
\begin{equation}
r{\cal R}=\left(1+{\alpha\over r}\right)^{(i\alpha\omega-s)/2}
\left(1-{\alpha\over r}\right)^{(i\alpha\omega+s)/2}
P_\nu^{(i\alpha\omega+s)/2,(i\alpha\omega-s)/2}(\alpha/r)
\end{equation}
This picture shows in particular how any given perturbation
leaves the future horizon of any observer
due to the representation as a wave in the $r^*$ and $t$
coordinate system. Therefore a set of perturbations that
where sufficiently well controlled in $\omega$ and angular
momentum would always die away from the local region of an
observer.

\section{Conclusions}

It is quite clear that non--abelian gauge theories lead to violations
of the no--hair theorems, but in all of the examples produced so far
these violations are not very severe. Most importantly of all, the
basic tenets of black hole thermodynamics still hold good and the
simplicity of the structure of the event horizon is maintained. This
shows up clearly in the formula for the action, which depends only on
quantities measured either at the event horizon or at infinity, and
suggests that the solutions are unique given the values of the angular
momentum $J$, the mass $M$ and {\it both} of the charges $Q_h$ and
$Q_\infty$.

Cosmological `no hair' theorems are important in the inflationary
universe scenario. The stability analysis of black holes solutions
gives us a way to approach the evolution of the early universe
and to address questions that involve large inhomogeneities, which
may have been present at the origin of our universe.

The conclusions reached from the study of black holes in de Sitter
space present far more of a challange to general relativity. The
existence of naked singularities in a spacetime with reasonable
matter, if we accept the cosmological constant, gives us a
good reason for studying the problems of quantum gravity.

\section{List of Symbols}

\begin{tabbing}

$A\cdot B=h^{ij}A_iB_j$ \qquad\qquad
$h_{ij}$=spatial metric\\
$A\wedge B=\epsilon^{ijk}A_jB_k$\qquad\quad~
$d\mu({\cal M})=$volume measure on ${\cal M}$\\
$\hbar$=Planck constant\qquad\quad
$S$=entropy \\
$c$=velocity of light\qquad\quad~
$\beta$=inverse temperature \\
$G$=Newton constant\qquad\quad
${\cal A}$=horizon area
\end{tabbing}

\section{References}

\def\prd#1,{{\it Phys. Rev. }{\bf D#1}}
\def\prl#1,{{\it Phys. Rev. Lett. }{\bf #1}}
\def\nuc#1,{{\it Nucl. Phys. }{\bf B#1}}

\re
Arnowitt, R., Deser S. and Misner, C. W., 1962, in {\it Gravitation, an
Introduction to Current Research} (Wiley, New York).
\re
Bardeen J. M., Carter B. and Hawking S. W., 1973, {\it Commun. Math.
Phys.} {\bf 31} 161
\re
Beckenstein J. D., 1972, {\it Lett. Nuovo. Cim.}, {\bf 4} 737.
\re
Bizon P., 1990, \prl64, 2844.
\re
Brady P. R. and Poisson E., 1992, {\it Class. Quantum Grav.} {\bf 9},
121.
\re
Brady P. R., Nunez D. and Sinha S.,1993, \prd47, 4239.
\re
Breitenlohner P., Forg\'acs P. and Maison D., 1992, \nuc383, 357.
\re
Brill, D. R., Horowitz, G. T., Kastor D. and Traschen, J., 1993,
preprint.
\re
Brown, J. D., Martinez E. A. and York J. W., 1991, \prl66, 2281.
\re
Brown, J. D., Martinez E. A. and York J. W., 1991, Ann. N.Y. Acad. Sci
{\bf 631}, 255.
\re
Carter B., 1968, Comm. Math. Phys. 10, 280.
\re
Carter B., 1973, in Black Holes ed C. and B. S. DeWitt
(New York, Gordon and Breach).
\re
Carter B., 1979, in {\it General Relativity: An Einstein Centenary
Survey,} edited by Hawking S. W. and Israel W. (C. U. P., Cambridge).
\re
Chambers C. and Moss I. G., 1993, Newcastle university preprint.
\re
Coleman S., Preskill J. and Wilczek F., 1991, \prl67, 1975
\re
Dowker F., Gregory R. and Traschen J., 1992, \prd45, 2762.
\re
Droz S., Heusler M. and Straumann N., 1991, {\it Phys. Lett. } {\bf
268}, 371.
\re
Gibbons, G. W. and Hawking S. W., 1977, \prd15, 2752.
\re
Gibbons, G. W. and Maeda K., 1988, \nuc298, 741.
\re
Hawking S. W., 1972, {\it Commun. Math. Phys.}, {\bf 25} 152.
\re
Hawking S. W., 1979, in {\it General Relativity: An Einstein Centenary
Survey,} edited by Hawking S. W. and Israel W. (Cambridge University
Press, Cambridge).
\re
Hawking S. W. and Moss I. G., 1982, {\it Phys. Lett. } {\bf 110B}
35.
\re
Israel, W., 1967, \prd164, 1776.
\re
Israel, W., 1992, in Black Hole Physics, ed De Sabbata V. and Zhang Z.
(Amsterdam: Kluwer Academic Publishers).
\re
Jacobson T., Kang G. and Myers R., preprint.
\re
K\"unzle H. P. and Masoud--ul--Alam A. K., 1990, {\it J. Math. Phys.}
{\bf 31} 928.
\re
Lee K., Nair V. P. and Weinberg E. J., 1992, \prd45, 2751.
\re
Luckock H. and I. G. Moss, 1986, {\it Phys. Lett.} {\bf B176} 341
\re
Luckock H., 1987, `Black hole Skyrmions' in String theory,
quantum cosmology and quantum gravity, eds. H. J. de Vega and N.
Sanchez, (World Scientific, Singapore).
\re
Maeda K., Tachizawa T., Torii T. and Maki T., 1993, Waseda University
preprint.
\re
Mellor F. and Moss I. G., 1990, \prd41, 403.
\re
Mellor F. and Moss I. G., 1989, {\it Class. Quantum Grav.} {\bf 6},
1379.
\re
Mellor F. and Moss I. G., 1992, {\it Class. Quantum Grav.} {\bf 9},
L43.
\re
Moss I. G., 1992, \prl69, 1852.
\re
Price R. H., 1972, \prd5, 2439.
\re
Ruffini, R and Wheeler, J. A., 1971, `Relatavistic cosmology and space
platforms' in {\it Proceedings of Conference on Space Physics} (ESRO,
Paris).
\re
Shiromizu T., Nakao K., Kodama H. and Maeda K., 1993, \prd47, R3099.
\re
Starobinsky A. A. and Churilov S. M., 1973, Soviet Phys. JETP 38,
1.
\re
Straumann N. and Zhou, Z. H., 1990, {\it Phys. Lett.} {\bf 237B}, 353
\re
Teukolsky S. A., 1972, \prl29, 1114.
\re
Visser M., 1993, Washington University preprint.
\re
Volkov M. S. and Gal'tsov D. V., 1989, J.E.T.P. Lett. {\bf 50} 346.
\re
Wald R. M., 1993, \prd48, R3427.

\end{document}